\documentclass[cmt,numbook]{svjour}
\usepackage{latexsym}
\usepackage{amssymb,amsmath}
\usepackage{graphicx}
\newcommand{\xx}{\mbox{\boldmath$x$}}

\newcommand{\vv}{\mbox{\boldmath$v$}}

\newcommand{\re}{r_e}

\newcommand{\Zq}{N_{q}}

\newcommand{\bftau}{\mbox{\boldmath$\tau$}}
\newcommand{\Tphys}{T_{\rm phys}}
\newcommand{\Cv}{C_{\rm\scriptscriptstyle V}}

\newcommand{\ue}{u_e}
\newcommand{\ve}{v_e}

\newcommand{\Dcrit}{D_{\rm crit}}
\begin{document}
%
\title{Self-gravitating Stellar Systems and Non-extensive Thermostatistics}
\subtitle{}
\author{Masa-aki Sakagami\inst{1} \and Atsushi Taruya\inst{2}
}                     
%
\mail{M. Sakagami (\email{sakagami@phys.h.kyoto-u.ac.jp})\
A. Taruya (\email{ataruya@utap.phys.s.u-tokyo.ac.jp})}
\institute{Graduate School of Human and Environmental Studies,
Kyoto University, Kyoto 606-8501, Japan 
\and Research Center for the Early Universe(RESCEU), 
School of Science, University of Tokyo, Tokyo 113-0033, Japan}
%
\date{Received: date / Accepted: date}
%
\maketitle
%
\begin{abstract}
After introducing the fundamental properties of self-gravitating systems,
we present an application of Tsallis' generalized entropy to the analysis of 
their thermodynamic nature. 
By extremizing the Tsallis entropy, we obtain an equation of 
state known as the {\it stellar polytrope}. 
For a self-gravitating stellar system confined within a perfectly
reflecting wall, we discuss
the thermodynamic instability caused by its negative specific heat. 
The role of the extremum as a quasi-equilibrium is also demonstrated 
from the results of $N$-body simulations.
\end{abstract}
%
%
%
%
%
%
\section{Introduction}
\label{intro}
In any subject of astrophysics and cosmology, many-body 
gravitating systems play an essential role. 
Globular clusters and elliptical galaxies, which are recognized as
self-gravitating stellar systems,  are 
typical examples \cite{BT1987,SP1987,EHI1987,MH1997}. 
Several beautiful works on the
thermodynamics of self-gravitating systems \cite{Antonov1962,LW1968} 
have shown  peculiar features such as a negative specific heat 
and the absence of global entropy maxima, 
which is referred to as {\it the gravothermal catastrophe}. (For a general 
introduction to this subject, see \cite{Padmanabhan1990}.)
Furthermore, the long-range nature of the gravitational interaction
makes a  discussion of the relationship between 
non-extensive thermostatistics the self-gravitating systems tempting.

Recently, a new framework for thermodynamics based on Tsallis' non-extensive 
entropy was proposed \cite{T1988}. It has been applied  extensively 
to deal with a variety of interesting problems to which standard
Boltzmann-Gibbs statistical mechanics can not be applied 
\cite{AO2001,KL2002}. The study of self-gravitating stellar systems has been
one of the most interesting applications of Tsallis' framework of
thermostatistics (see, e.g., \cite{PP1993,A1993,LSS2002}). 
Here, some of the  progress 
in its application to  stellar systems \cite{TS2002a,TS2002b,TS2003,TSPRL} 
will be reviewed.   
Although its dynamics is complicated in general, 
if we impose spherical symmetry on the system, its treatment is
considerably simplified due to the well-known $1/r^2$ behavior of 
the gravitational force. Furthermore the spherical system still keeps
the remarkable nature of a negative specific heat \cite{Antonov1962,LW1968}. 
Thus, self-gravitating stellar systems seem to be a desirable testing ground 
for Tsallis' non-extensive thermostatistics.

This paper is organized as follows. In section 2, we briefly review  
the principal characteristics of self-gravitating systems. In particular, 
the standard treatment for 
the gravothermal catastrophe based on the Boltzmann--Gibbs entropy is explained.
Then, its extension to the Tsallis entropy is discussed in section 3. 
The properties of  states of the stellar system that extremize the Tsallis 
entropy are clarified. In section 4, the interesting  role of the above states 
as quasi-equilibria is discussed from the results of numerical simulations.       
Finally, section 5 is devoted to discussion and conclusions.
%
%
%
%
%
%
\section{Some basic properties of the self-gravitating systems}
\label{basic}
%
%
%
%
%
%
In order to get an intuitive understanding of the evolution of 
many-body self-gravitating systems, let us consider a very simple 
situation: the circular motion of a particle of  mass $m$ and 
 velocity $v$ at a radius $r$ in a spherical mass distribution 
with a constant density $\rho_0$:
\begin{equation}
m\frac{v^2}{r} = \frac{G m M(r)}{r^2},~~~M(r) = \frac{4\pi}{3}r^3\rho_0,
\label{eq:circular}
\end{equation}
where $G$ is the gravitational constant and $M(r)$ is the mass contained 
within the sphere of  radius $r$.
By means of the orbital period $T$, we define the dynamical time of 
the system (see, e.g., p.57 of \cite{BT1987}) as 
\begin{equation}
t_{\rm dyn} = \frac{T}{4} = \sqrt{\frac{3\pi}{16 G\rho_0}},
\label{eq:dtime}
\end{equation} 
which is recognized as the characteristic time scale of more general 
self-gravitating systems with mean density $\rho_0$. 
A distribution of the self-gravitating system evolves with the time scale
\eqref{eq:dtime}, so this quantity is called the dynamical time. As explained in the
Appendix, the many-body self-gravitating system has another timescale, 
i.e. the relaxation time \cite{Chandra} (see also chapter 4 of  
\cite{BT1987}), 
\begin{equation}
t_{\rm rel} \sim \frac{0.1 N}{\ln N} t_{\rm dyn}
\label{eq:rtime}
\end{equation}
where $N$ is the number of particles in the system.  This time scale 
represents the relaxation processes due to scattering by the gravitational
interaction  between each pair of particles. 

In a usual system, we believe that a distribution of particles
approaches the so-called thermal equilibrium state 
within the relaxation time \eqref{eq:rtime}.
However, this is not necessarily the case for self-gravitating systems,
since they have the peculiar property of a {\it  negative specific heat},
which is explained through the following discussion. For the circular 
orbit obeying \eqref{eq:circular}, the virial theorem is easily
derived as 
\begin{equation}
2 K + U = 0, 
\label{eq:virial}
\end{equation} 
where $K = m v^2 /2$ is the kinetic  and 
$U = - GmM(r)/r$ the potential energy. 
In general, for self-gravitating many-body systems consisting of 
$N$ particles, a virial relation identical to \eqref{eq:virial} holds 
between  long-time-averaged values of
the total kinetic and the total potential energy 
(see, e.g., chapter 8.1 of \cite{BT1987}).
Thus, the total energy $E$ can be expressed as 
\begin{equation}
E = K + U = -K .
\label{eq:negative}  
\end{equation}
If we introduce a temperature to characterize the total kinetic 
energy as $K= 3NkT/2$, where $k$ is the Boltzmann constant,
then \eqref{eq:negative} tells us that 
the specific heat of a self-gravitating system is {\it negative}, since
\begin{equation}
C \equiv \frac{dE}{dT} = - \frac{3Nk}{2} < 0 ,
\label{eq:naive}
\end{equation}
which suggests the existence of a thermodynamic instability
proceeding with the relaxation timescale \eqref{eq:rtime}.

For a more precise discussion of the thermodynamic properties of
self-gravitating systems, Antonov \cite{Antonov1962} and Lynden-Bell 
and Wood \cite{LW1968} considered a system confined 
within a spherical adiabatic wall of  radius $r_e$. 
This means that the $N$ particles 
in this system interact via Newtonian gravity and bounce elastically from 
the wall. In order to investigate the equilibrium states of the system,
the Boltzmann--Gibbs entropy of the system,      
\begin{equation}
  \label{eq: BG_entropy}
  S_{\rm BG}=-\int f\,\ln\,f\,\,d^6\bftau, 
\end{equation}
is introduced, where $f(\xx,\vv)$ is the distribution 
function in phase-space. Here, the phase space measure is defined as
\begin{equation}
d^6\bftau \equiv \frac{d^3\xx d^3\vv}{h^3},~~~h^3 \equiv (l_0v_0)^3 
\label{eq:measure}
\end{equation}
where $h^3$  denotes the phase-space element with unit length
$l_0$ and unit velocity $v_0$. 
Hereafter, we set the Boltzmann constant to be unity, i.e. $k=1$.
Under the condition that the total mass $M$ and the energy $E$ of 
the system are kept constant, we maximize the entropy $S_{\rm BG}$:
\begin{equation}
\delta S_{\rm BG} - \alpha \delta M - \beta \delta E = 0,
\end{equation}  
where $\alpha$ and $\beta$ are Lagrange multipliers.
Through this procedure, the equation of state of the system 
turns to be {\it isothermal} \cite{LW1968}:
\begin{eqnarray}
p(\xx) &\equiv & \int \frac{1}{3} v^2 f\,\, \frac{d^3\vv}{h^3} = 
\frac{\rho(\xx)}{\beta},\cr
\rho(\xx) &\equiv & \int f\,\, \frac{d^3\vv}{h^3}
\label{eq: EOS_iso}
\end{eqnarray}
where $p$ and $\rho$ are the pressure and density, respectively.
The thermal equilibrium states of ordinary systems are homogeneous; 
i.e. they have  uniform density. 
However, this is not the case for self-gravitating systems. 
For the spherical systems we are discussing here, the magnitude of the density and 
pressure should increase as the radius $r$ decreases in order to support 
the system against its self-gravity. 
It is shown that a series of equilibrium states 
is well parameterized by the value of the density contrast $D = \rho_c/\rho_e$, 
where $\rho_c$ is the central density and $\rho_e = \rho(r_e)$ is the density 
at the wall.  In the next section, 
we will present the procedures for calculating the equilibrium state 
in more general situations.   
\begin{figure}[t]
\begin{minipage}{\textwidth}
   \includegraphics*[width=7.5cm]{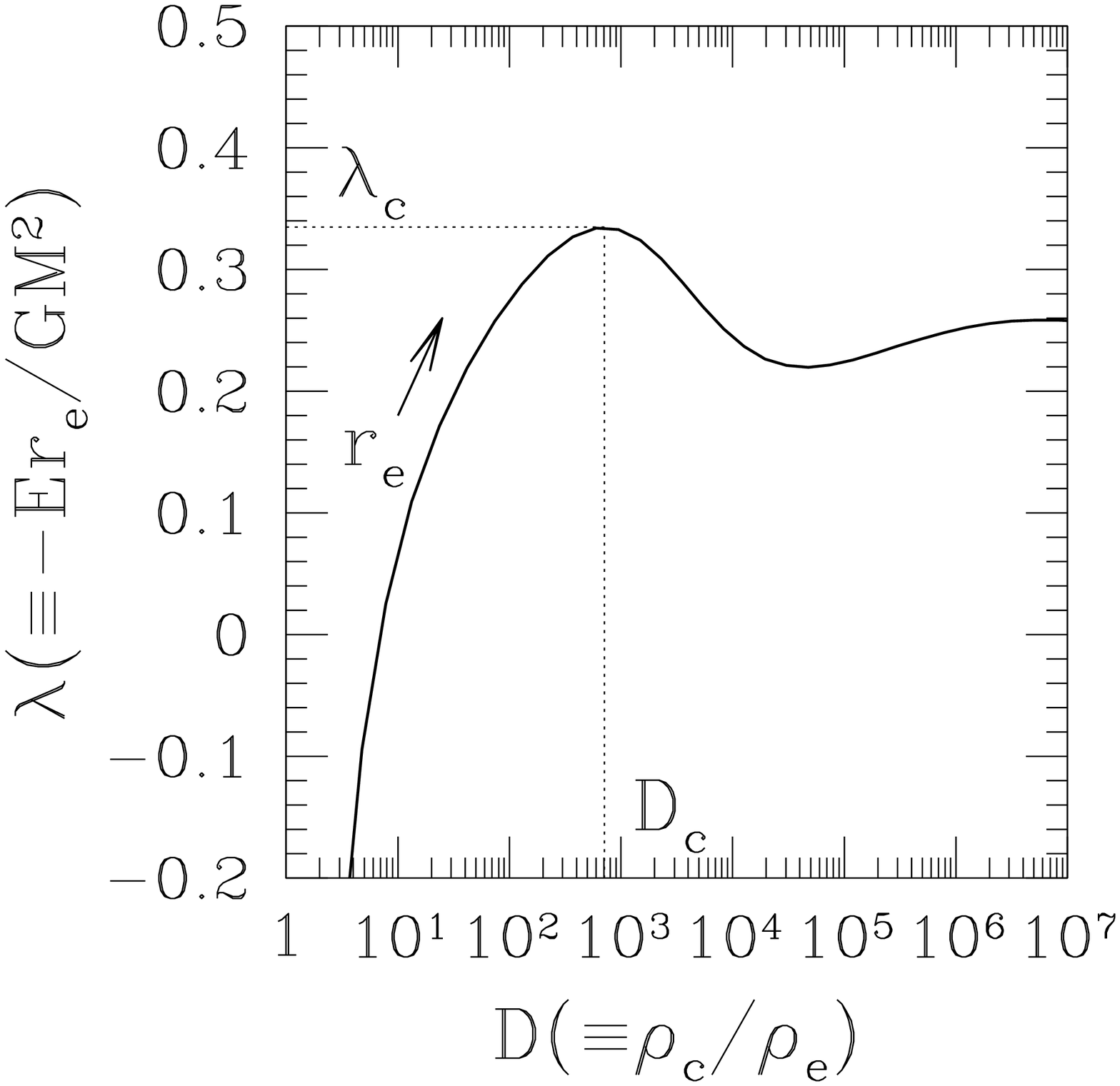}
   \hfill
   \parbox[b]{8.cm}{
   \includegraphics*[width=7.5cm,height=7.5cm]{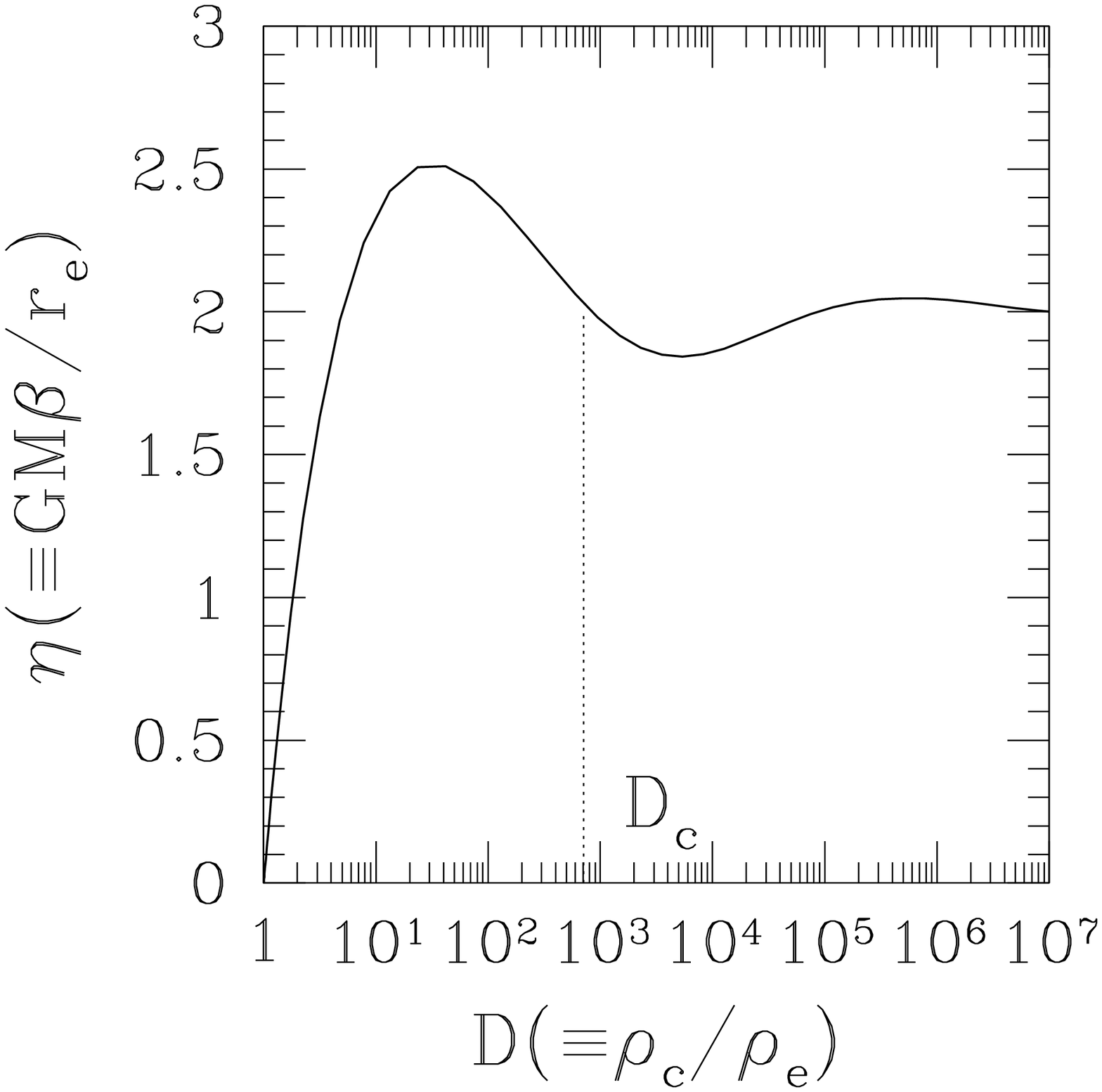}
   \hfill
   }
\end{minipage}
\caption[]{Energy-radius--density contrast (left) and
radius-temperature--density contrast (right) relationships for the isothermal
stellar system}
\label{fig: turning_iso}
\end{figure}%
In Fig.~\ref{fig: turning_iso} (left), we plot the dimensionless quantity 
$\lambda = - r_e E /GM^2$ as a function of the density contrast $D$.
Since the solid curve in this figure represents states which extremize
the entropy, we note that no equilibrium state is attained when 
the radius of the wall $r_e$ is larger than the critical radius
\begin{equation}
r_c = 0.335 \frac{GM^2}{(-E)}, 
\end{equation}   
which corresponds to the state with the critical density contrast
$D_c = 709$. 
Furthermore, it can be shown, from a turning-point analysis
for a linear series of equilibria
\cite{LW1968,Padmanabhan1989,Katz1978,Katz1979}, 
that, along the curve $\lambda(D)$ derived 
from the condition $\delta S_{\rm BG} = 0$,
all states with $D >  D_c$ are unstable. 
Also, the explicit evaluation of the eigenmodes for the second variation of 
the entropy $\delta^2 S_{\rm BG}$ leads to the same results 
\cite{Padmanabhan1990,Padmanabhan1989}.
We call this phenomenon, i.e. the absence of thermal equilibrium in 
self-gravitating systems when $r > r_c $ and/or $D > D_c$,  
{\it the gravothermal catastrophe}.

Heuristically, this instability is 
explained by the presence of the negative specific heat as follows. 
In a fully relaxed gravitating system with a sufficiently larger radius, 
the negative specific heat arises at the inner part of the system,  
where we have $C_{\rm V,inner}<0$, 
while the specific heat at the outer part remains positive 
($C_{\rm V,outer}>0$), since one can safely neglect the effect of self-gravity. 
In this situation, if a tiny heat flow is momentarily supplied from the inner 
to the outer part,  both the inner and outer parts get hotter after the 
readjustment of the system. Now imagine the case 
$C_{\rm V, outer}>|C_{\rm V, inner}|$.  The outer part has so much thermal 
inertia that it cannot heat up as fast as the inner part, and thereby the 
temperature difference between inner and outer parts increases. As a 
consequence, the heat flow never stops, leading to a catastrophic 
temperature growth.

\begin{figure}
\begin{minipage}{\textwidth}
   \includegraphics*[width=7.5cm,height=7cm]{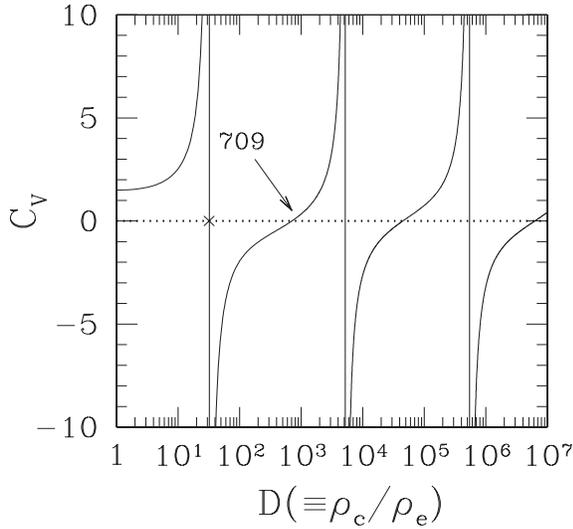}
   \hfill
   \parbox[b]{7.5cm}
         {\caption[]{Specific heat as a function of the density contrast 
$D = \rho_c/\rho_e$ for the isotropic stellar system.}
          \label{fig: Cv_iso}}
\end{minipage}
\end{figure}

Fig. \ref{fig: turning_iso}(right) plots the dimensionless 
inverse temperature
$\eta \equiv GM\beta/r_e$ with respect to the density contrast.
We can evaluate the specific heat at constant volume,
\begin{equation}
C_{V} \equiv \left(\frac{dE}{dT}\right)_{r_e} 
      =  - \beta^2 \left(\frac{dE}{d\beta}\right)_{r_e}
      =  M \eta^2 \dfrac{\left(\dfrac{d \lambda}{d D}\right)_{r_e}}
                   {\left(\dfrac{d \eta}{d D}\right)_{r_e}}~~,
\label{eq:c_v}
\end{equation}
as shown in Fig.\ref{fig: Cv_iso}. 

In contrast to the naive discussion for 
\eqref{eq:naive} based on the virial theorem, the actual specific heat 
\eqref{eq:c_v} changes sign many times. We observe that 
a self-gravitating system confined by a wall of smaller radius
has a positive specific heat. The reason is as follows.
It can be shown that the virial theorem 
for a system surrounded by a boundary leads not to 
\eqref{eq:virial} but to
\begin{equation}
2 K + U = 3 P_e V ,
\label{eq:virial_wall}
\end{equation}
where $P_e$ is the pressure at the wall and $V$ 
the volume of the system (see, e.g.,  eq.(29) of \cite{TS2002a}). 
For the case of a smaller radius,
the surface energy dominates the gravitational energy of
the bulk so that the system behaves as a normal one; i.e. $C_V > 0$.
As the radius $r_e$ increases, the gravitational energy becomes
significant. Eventually, the specific heat changes its sign, $C_V < 0$ ,
when the value of the density contrast becomes $D = 32.2$, which corresponds 
to a peak of the $\eta(D)$ curve (the left side of Fig.\ref{fig: turning_iso} ).

At the onset of the gravothermal catastrophe, $D_c = 709$,
the specific heat vanishes, and $C_V > 0$ in the unstable 
segment of the $\lambda (D)$ curve, $D > 709$. This fact implicitly 
indicates a role of the negative specific heat in existence of
the thermodynamic instability in  self-gravitating systems.
Following the previous heuristic argument, above the critical point
($D > D_c = 709$), the specific heat of the total system becomes positive
($C_{\rm V} = C_{\rm V, inner} + C_{\rm V, outer} > 0$) equivalently,
\begin{equation}
C_{\rm V, outer} > -\, C_{\rm V, inner} > 0 ,
\label{eq: C-outer-inner}
\end{equation}
which means the outer normal part of the system has
a larger heat capacity, so that it cannot catch up with the
temperature increase of the inner part.

So far, we have discussed some interesting features of  self-gravitating 
systems. Among these properties, the peculiarity of the system arising from 
the negative specific heat has been studied in terms of 
the Boltzmann--Gibbs entropy \eqref{eq: BG_entropy}. 
In the next section, this sort of the analysis will be extended to 
that based on the non-additive entropy, i.e. the Tsallis entropy.
However, before closing this section, we briefly summarize 
the subsequent investigation of the gravothermal catastrophe.

In this paper we adopt the thermodynamic approach in terms of 
the entropy and/or the free energy in order to 
investigate instabilities of  self-gravitating systems.  
Although this method has the advantage of simplicity, 
we cannot deal with dynamical aspects of the instability 
appearing within the relaxation timescale \eqref{eq:rtime}. 
Typically, each particle (star) in a self-gravitating system 
has a long mean free path compared  with 
its system size.  In other words, during its travel across the system,
the the motion of a particle is driven by the gravitational potential of the whole 
system and the particle experiences quite a few two-body encounters,
so that the relaxation time $t_{\rm rel}$ \eqref{eq:rtime} is  
much longer than the dynamical one $t_{\rm dyn}$ \eqref{eq:dtime}.  
This means that we cannot rely on the usual notion of  local equilibrium,  
which is realized in  cases with a short mean free path. 
Thus, 
the Fokker-Planck equation (see, e.g., chapter 8.3 of \cite{BT1987}),
which can be derived from the expansion of a collision integral in 
the Boltzmann equation with respect to the 
momentum transfer due to a two-body collision, is a useful 
approach for studying the long-term evolution of a self-gravitating system.
By means of the linearized Fokker-Planck equation, 
eigenvalues and eigenfunctions which describe 
the time evolution of perturbations around the thermal equilibrium
have been obtained \cite{Inagaki1980}.

The gas model in which a star (particle) in the 
self-gravitating many-body system 
is replaced by a gas particle, 
is an alternative approach. Although the gas model loses some of 
the basic properties of a the self-gravitating many-body system, 
we note that it still shows the interesting aspects of 
the gravothermal catastrophe. In this model, the local relaxation time
is shorter than the dynamical timescale. Thus, we can introduce the
usual notions of  local equilibrium and the local temperature, 
which make our analysis much easier than in the stellar system. 
The local specific heat is evaluated to show that the condition 
\eqref{eq: C-outer-inner} characterizes the onset of 
the gravothermal catastrophe \cite{HS1978}.
The time evolution of a linear perturbation around 
the equilibrium gas with heat transfer has been investigated to show that
the proposed mechanism for the instability based on the negative specific heat
actually works in self-gravitating gaseous systems \cite{MH1991}.  
%
%
%
%
%
%
%
%
%
\section{Analysis of gravothermal catastrophe based on Tsallis' 
generalized entropy}
\label{sec:3}
%
%
%
%
%
\subsection{Tsallis entropy and stellar polytrope}
\label{sec:3.1}

Here we shall discuss a generalization of the analysis in 
previous section based on the Boltzmann--Gibbs entropy \eqref{eq: BG_entropy}
to the case of the Tsallis entropy.
The details are given in the literature \cite{TS2002a,TS2002b,TS2003}. 
The the energy and mass of a  one- particle distribution $f(\xx,\vv)$ 
are respectively expressed as  
\begin{eqnarray}
&&   E= K+U = \int \left\{ \frac{1}{2}\,v^2 + \frac{1}{2}\,
  \Phi(\xx) \right\}\,\,f(\xx,\vv)\,\,d^6\bftau, 
\label{eq: def_E} \\
&&    M = \int \,\, f(\xx,\vv)\,\,d^6\bftau, 
\label{eq: def_M}
\end{eqnarray}
where the quantity $\Phi(\xx)$ is the gravitational potential 
given by 
\begin{eqnarray}
  \Phi(\xx) = -G\int \frac{f(\xx',\vv')}{|\xx-\xx'|}\,d^6\bftau'.
\label{eq: def_Phi}
\end{eqnarray}

As for generalization, the most crucial problem is the choice of the 
statistical average in non-extensive thermostatistics. 
Following a seminal paper \cite{PP1993}, 
the analyses in a couple of our papers \cite{TS2002a,TS2002b} 
have been done by utilizing the {\it old} Tsallis formalism with the 
standard linear mean values(see also \cite{Chavanis,CS2003} for 
comparative works). On the other hand, 
a more sophisticated framework by means of 
the normalized $q$-expectation values has recently been presented 
\cite{TMP1998,MNPP2000}. As several authors have 
advocated, the analysis using normalized $q$-expectation values is thought to be 
essential, since the undesirable divergences in some physical systems 
can be eliminated safely when introducing the normalized $q$-expectation values. 
Furthermore, non-uniqueness of the Boltzmann-Gibbs theory has been 
shown by using normalized $q$-expectation values \cite{AR2000}. 
In this paper, we mainly report our investigation \cite{TS2003} based on 
normalized $q$-expectation values. However, this does not imply that 
all the analyses with standard linear means or un-normalized $q$-expectation values 
lose the physical significance. Actually, we observed 
some differences between results based on two frameworks, which will be 
mentioned at the end of this section.

In the new framework of Tsallis' non-extensive 
thermostatistics, all the macroscopic observables of the quasi-equilibrium 
system can be characterized by the escort distribution, but the 
escort distribution itself is not thought to be fundamental. 
Rather, there exists a more fundamental probability function 
$p(\xx,\vv)$ that quantifies the phase-space structure. 
With the help of this function, the escort distribution is defined and 
the macroscopic observables are expressed as  normalized 
$q$-expectation value as follows (see, e.g., \cite{TMP1998,MNPP2000}): 
\begin{eqnarray}
&  \mbox{escort distribution}~~ &: 
P_q(\xx,\vv)= \frac{\displaystyle \left\{p(\xx,\vv)\right\}^q}
  {\displaystyle \int d^6\bftau \left\{p(\xx,\vv)\right\}^q}\,\,\,, 
\label{eq: def_of_escort}   
\\
&  \mbox{normalized q-value}~~ &: 
\langle {\it O}_i\rangle_q = \int d^6 \bftau \,\,{\it O}_i\,\,
P_q(\xx,\vv)\,\,.
\label{eq: def_of_q-values} 
\end{eqnarray}
Based on the fundamental probability $p(\xx,\vv)$, 
the Tsallis entropy is given by
\begin{equation}
        S_q=-\frac{1}{q-1}\,\int d^6\bftau \,
\left[\left\{p(\xx,\vv)\right\}^q-p(\xx,\vv)\right].
\label{eq: Tsallis entropy}
\end{equation}
Note that the probability $p(\xx,\vv)$ satisfies the 
normalization condition:  
\begin{equation}
  \label{eq: normalization_p(x,v)}
  \int\,\,d^6\bftau\,\,p(\xx,\vv) = 1.
\end{equation}

To  apply the above Tsallis formalism to the present problem 
without changing the definition of energy and mass 
(\ref{eq: def_E}) and (\ref{eq: def_M}), we identify the 
one-particle distribution with the escort distribution $P_q$, not 
the probability function $p(\xx,\vv)$:  
\begin{equation}
  \label{eq: escort_dist}
  f(\xx,\vv) = M\,\,\frac{\{p(\xx,\vv)\}^q}{\Zq}\,\,; 
\,\,\,\,\,\,\,\,\,\,\,\,
  \Zq = \int d^6\bftau\,\,\{p(\xx,\vv)\}^q
\end{equation}
so as to satisfy  mass conservation (\ref{eq: def_M}). 
The definitions of the Tsallis entropy 
\eqref{eq: Tsallis entropy} and the escort distribution 
\eqref{eq: escort_dist} clearly show that the Boltzmann-Gibbs entropy
\eqref{eq: BG_entropy} is recovered in the limit $q \rightarrow 1$.
Now, adopting the relation (\ref{eq: escort_dist}), let us seek the 
extremum-entropy state under the constraints (\ref{eq: def_E}) and 
(\ref{eq: normalization_p(x,v)}).
The variational problem is  
\begin{equation}
 \delta\left[ S_q- \alpha\left\{\int d^6\bftau p-1\right\}-
\beta\left\{ \int d^6\bftau 
\left(\frac{1}{2}v^2+\frac{1}{2}\Phi\right)f -  E\right\}\right] = 0,
\label{eq: Max.Ent.Prin}
\end{equation}
where the variables $\alpha$ and $\beta$ denote the Lagrange multipliers. 
The variation with respect to the probability $p(\xx,\vv)$ gives 
the   so-called {\it stellar polytrope} as the extremum-entropy state
\cite{TS2003} (see also p.223 of \cite{BT1987}):
\begin{equation}
f(x,v) = M\,\,\frac{\{p(x,v)\}^q}{\Zq}= 
A \left[\Phi_0 - \frac{1}{2}v^2 - \Phi(x) \right]^{q/(1-q)}, 
\label{eq: extremum_state}
\end{equation}
where we define the constants $A$ and $\Phi_0$ as  
\begin{eqnarray}
A = \frac{M}{\Zq}\,\, \left\{ \frac{q(1-q)}{\alpha(1-q)+1} 
        \frac{\beta M }{\Zq}\right\}^{q/(1-q)},~~~\Phi_0
 = \frac{\Zq}{\beta M(1-q)} + <\varepsilon>,~~~
\label{eq: A_Phi0}
\end{eqnarray}
with the quantity $<\varepsilon>$ being 
\begin{equation}
<\varepsilon> = \frac{1}{M}\,\int d^6\bftau 
\left(\frac{1}{2}v^2 + \Phi\right) f. 
\label{eq: epsilon}
\end{equation}

By means of the distribution function \eqref{eq: extremum_state},
we evaluate the density $\rho(r)$ and the 
isotropic pressure $P(r)$ at the radius $r=|\xx|$ as    
\begin{eqnarray}
\rho(r) &\equiv& \int\frac{d^3\vv}{h^3}\,\,\, f(\xx,\vv),
\nonumber \\
&=& 4\sqrt{2}\pi \,B\left(\frac{3}{2},\frac{1}{1-q}\right)
  \,\frac{A}{h^3}\,\,[\Phi_0-\Phi(r)]^{1/(1-q)+1/2}
\label{eq: density}
\end{eqnarray}
and 
\begin{eqnarray}
P(r) &\equiv& \int\frac{d^3\vv}{h^3} \,\,\,\frac{1}{3}\, v^2 \,
f(\xx,\vv),
\nonumber \\
&=& \frac{8\sqrt{2}\pi }{3}\,B\left(\frac{5}{2},\frac{1}{1-q}\right)
  \,\frac{A}{h^3}\,\,[\Phi_0-\Phi(r)]^{1/(1-q)+3/2}.  
\label{eq: pressure}
\end{eqnarray}
Here, the function $B(a,b)$ denotes the beta function. 
These two equations lead to the polytropic relation 
\begin{equation}
  \label{eq: polytrope}
  P(r) = K_n\rho ^{1+1/n}(r), 
\end{equation}
with the polytrope index $n$ related to the parameter $q$ by
\begin{equation}
  \label{eq: index}
  n = \frac{1}{1-q} \,+ \,\frac{1}{2}.
\end{equation}
The explicit form of the dimensional constant $K_n$ is 
given in \cite{TS2003}.  We note that the isothermal 
equation of state \eqref{eq: EOS_iso} discussed in the previous section 
arises in the limit $n \rightarrow \infty \,\, (q \rightarrow 1) $ as expected. 

In terms of $K_n$ and $\rho$, the one-particle distribution can be 
rewritten as  
\begin{eqnarray}
f(x,v)&=& \frac{1}{4\sqrt{2}\pi\,\,B(3/2,n-1/2)}\,\,\,
\frac{\rho\,\,h^3}{ \{ (n+1)\,K_n\,\rho^{1/n}\}^{3/2}}  
\nonumber \\
&&~~~~~~~~~~~~~~~~~~~~~~~~~~~~~~~~
\times\,\,\,\left\{1-\frac{v^2/2}{(n+1)\,K_n\,\rho^{1/n}}\right\}^{n-3/2},    
  \label{eq: poly_dist}
\end{eqnarray}
which agrees with the result based on the {\it old} Tsallis formalism.
That is to say, the equilibrium distribution 
(\ref{eq: poly_dist}) turns out to be invariant irrespective of the 
choice of the statistical averages.

We note that the equilibrium distribution (\ref{eq: extremum_state}) 
with (\ref{eq: A_Phi0}) contains the new quantities $\Zq$ and $<\varepsilon>$, 
which implicitly depend on the distribution function itself. 
In marked contrast to 
the result in the old Tsallis formalism,    
this fact gives rise to  non-trivial thermodynamic relations as follows. 
By using the definitions of density and pressure (\ref{eq: density}) and 
(\ref{eq: pressure}),  the quantity $<\varepsilon>$ becomes
\begin{eqnarray}
  <\varepsilon> &=& \frac{1}{M}\,
\left\{\frac{3}{2}\,\int d^3x\,P(x) + \int d^3x\,\rho(x)\,\Phi(x)\,\right\}
\nonumber \\
&=& \frac{1}{M}\,
 \left\{\frac{3}{2}\,\int d^3x\,P(x) 
        - \int d^3x\,\rho(x)\,[\Phi_0-\Phi(x)]\,\right\} + \Phi_0.
\nonumber
\end{eqnarray}
Furthermore, by using the relation $\Phi_0-\Phi(x)= (n+1)(P/\rho)$ from 
(\ref{eq: density}) and (\ref{eq: pressure}) and substituting the equation 
(\ref{eq: A_Phi0}) into the above expression,  the quantity $<\varepsilon>$ 
is canceled and the equation reduces to 
\begin{equation}
\frac{\Zq}{\beta}\,=\,\int d^3x\,P(x).
\label{eq: beta-P_relation}
\end{equation}
As for  $\Zq$, the normalization 
condition (\ref{eq: normalization_p(x,v)}) gives us its actual 
expression \cite{TS2003}.

\subsection{Gravothermal catastrophe in the stellar polytropic system }
\label{sec: Gravo_stellar}

We will specifically focus on the spherically symmetric case with the 
polytrope index $n>3/2\,(q>0)$, in which the equilibrium distribution 
is at least dynamically stable (see chapter 5 of \cite{BT1987}). 
In this case,  the stellar equilibrium distribution 
can be characterized by the so-called {\it Emden solutions} 
(see, e.g., \cite{Chandra1939,KW1990}) and all the physical 
quantities are expressed in terms of the homology invariant 
variables $(u,v)$, which are subsequently used in later analysis.

We note that the one-particle distribution function 
(\ref{eq: extremum_state}) does not yet completely specify 
the equilibrium configuration, 
due to the presence of the gravitational potential, which implicitly 
depends on the distribution function itself. Hence, we need to 
specify the gravitational potential or density profile.
From the gravitational potential (\ref{eq: def_Phi}), 
we obtain the Poisson equation 
\begin{equation}
\label{eq: poisson_eq}  
 \frac{1}{r^2}\frac{d}{dr}\left\{r^2\frac{d\Phi(r)}{dr}\right\}=
  4\pi G \rho(r).
\end{equation}
By combining the above equation with (\ref{eq: density}), we 
obtain the ordinary differential equation for $\Phi$. 
Alternatively, a set of equations which represent 
the hydrostatic equilibrium are derived by using 
(\ref{eq: poisson_eq}), (\ref{eq: density}), and (\ref{eq: pressure}):  
\begin{eqnarray}
& \dfrac{dP(r)}{dr}\,=&\,\,-\frac{Gm(r)}{r^2}\,\rho(r), 
\label{eq: hydro_1}
\\
& \dfrac{dm(r)}{dr}\,=&\,\,4\pi\rho(r)\,r^2.  
\label{eq: hydro_2}
\end{eqnarray}
The quantity $m(r)$ denotes the mass evaluated at the radius $r$ 
inside the wall. Denoting the central density and pressure by 
$\rho_c$ and $P_c$, we then introduce the dimensionless quantities: 
\begin{equation}
\label{eq: dimensionless}
 \rho=\rho_c\,\left[\theta(\xi)\right]^n,\,\,\,\,\,\,
r=\left\{\frac{(n+1)P_c}{4\pi G\rho_c^2}\right\}^{1/2}\,\xi, 
\end{equation}
which yields the ordinary differential equation 
\begin{equation}
 \theta''+\frac{2}{\xi}\theta'+\theta^n=0,
\label{eq: Lane-emden_eq}
\end{equation}
where the prime denotes the derivative with respect to $\xi$. 
The quantities $\rho_c$ and $P_c$ in (\ref{eq: dimensionless}) 
are the density and the pressure at $r=0$, respectively. 
To obtain the physically relevant solution of (\ref{eq: Lane-emden_eq}), 
we use the  boundary condition
\begin{equation}
 \theta(0)=1, \,\,\,\,\,\,\,\theta'(0)=0.    
\label{eq: boundary}
\end{equation}
A family of solutions satisfying (\ref{eq: boundary}) is referred to 
as the {\it Emden solution}, which is well-known in the subject of 
stellar structure (see, e.g., chapter IV of ref.\cite{Chandra1939}).  
To characterize the equilibrium properties of Emden solutions,  
it is convenient to introduce the following 
set of variables, referred to as homology invariants 
\cite{Chandra1939,KW1990}: 
\begin{eqnarray}
 u &\equiv& \frac{d\ln m(r)}{d\ln r}=
\frac{4\pi r^3\rho(r)}{m(r)}=-\frac{\xi\theta^n}{\theta'},
\label{eq: def_u}
\\
\nonumber\\
 v &\equiv&  - \frac{d\ln P(r)}{d\ln r}=
\frac{\rho(r)}{P(r)}\,\,\frac{Gm(r)}{r}
=-(n+1)\frac{\xi\theta'}{\theta},  
\label{eq: def_v}
\end{eqnarray}
which reduce the degree of (\ref{eq: Lane-emden_eq}) 
from two to one. Recall that we are investigating the thermodynamic 
properties of a self-gravitating system within a wall of 
radius $r_e$. We can evaluate the total energy of the confined  
stellar system in terms of the pressure $P_e$, the density 
$\rho_e$ at the boundary $\re$, and the total mass $M$: 
\begin{eqnarray}
  E= K+U &=& \frac{3}{2}\,\int_{0}^{\re}dr\,4\pi r^2\,P(r) 
  -\int_0^{\re}\,dt\,\frac{Gm(r)}{r}\,\frac{dm}{dr}
\nonumber \\
&=& -\frac{1}{n-5}\,
\left[\,\frac{3}{2}\left\{ \frac{GM^2}{\re}-(n+1)\frac{MP_e}{\rho_e}\right\}
+(n-2)\,4\pi\,\re^3\,P_e\,\right], 
\nonumber 
\end{eqnarray}
by which the dimensionless quantity $\lambda$ can be 
expressed as a function of the homology invariants at the wall
 \cite{TS2002a,TS2002b,TS2003} : 
\begin{equation}
\lambda \equiv - \dfrac{E r_e}{GM^2} = - \frac{1}{n-5}\,
\left[\,\frac{3}{2}\left\{1-\frac{n+1}{\ve}\right\}+(n-2)\frac{\ue}{\ve}
\,\right].
  \label{eq: energy_uv}
\end{equation}
\begin{figure}
\begin{minipage}{\textwidth}
   \includegraphics*[width=11cm,height=6cm]{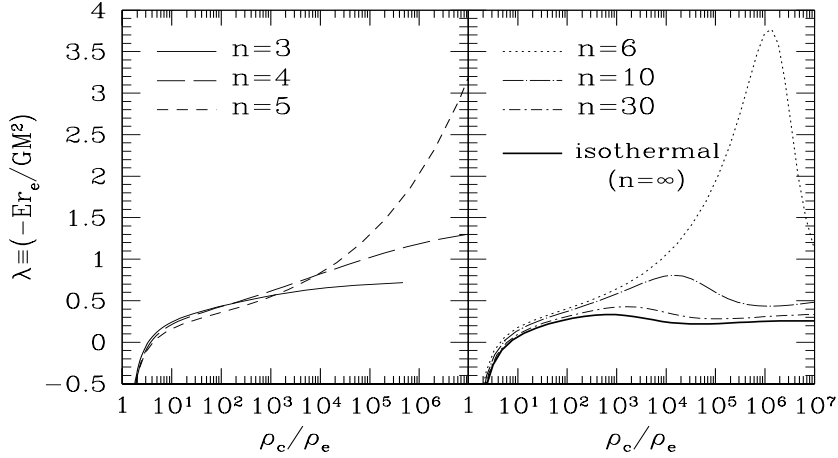}
   \hfill
   \parbox[b]{4cm}
         {\caption[]{Energy-radius-density contrast relationship for
                     the stellar polytropes with various indices. }
          \label{fig: lambda_poly}}
\end{minipage}
\end{figure}%
As was already shown in section \ref{basic} for the isothermal case,
the above dimensionless quantity $\lambda$ has an important role
for the analysis of the gravothermal catastrophe.
Figure \ref{fig: lambda_poly} shows $\lambda$ 
as a function of the ratio of the central density to that at the 
boundary, $\rho_c/\rho_e$. 
We notice that the $\lambda$-curves are bounded from above and have peaks in 
the case of $n>5$ ({\it right panel}).
On the  other hand, the curves for $n\leq5$ monotonically increase
({\it left panel}).   
It follows that the stellar polytrope within an adiabatic wall 
exhibits the gravothermal instability in the case of a polytropic 
index $n > 5$. 
Similarly to the isothermal case, the evaluation of eigenvalues for 
the second variation of the Tsallis entropy $\delta^2 S_q $ gives the   
same results as the above turning point analysis in terms of 
the $\lambda$-curve \cite{TS2002a,TS2002b,TS2003}.

\subsection{Thermodynamic instability arising from the negative 
specific heat}
\label{sec:3.3}

As discussed in the isothermal case in the previous section, 
thermodynamic instability in stellar systems is 
intimately related to the presence of a negative specific heat 
\cite{LW1968,LyndenBell1999}. 
The evaluation of the specific heat is thus necessary for clarifying the 
thermodynamic properties. In this regard, the identification of the
temperature in the stellar system in the polytropic case 
is the most essential task.  

In the new framework of Tsallis' non-extensive thermostatistics, 
the physically plausible thermodynamic temperature $\Tphys$ can be 
defined from the zero-th law of thermodynamics 
\cite{AMPP2001,Abe2001,MPP2001}. The thermodynamic 
temperature in a non-extensive system differs from 
the usual one, i.e., the inverse of the Lagrange multiplier $\beta$. 
The pseudo-additivity of the Tsallis entropy and the transitivity of 
thermal equilibria suggests 
\begin{equation}
\Tphys=
\,[1+(1-q)\,S_q]\,\,\beta^{-1}  . 
\label{eq: def_Tphys}
\end{equation}
Here, the relation (\ref{eq: def_Tphys}) 
should be carefully applied to the present case, since the  
verification of zero-th thermodynamic law is very difficult 
in a stellar equilibrium system with long-range interactions. 
Furthermore, even using the new formalism, the energy $E$ 
remains non-extensive due to the self-referential form of the 
potential energy (see (\ref{eq: def_E}) and (\ref{eq: def_Phi})). 
However, the consistency of the physical temperature given by 
(\ref{eq: def_Tphys}) can be shown by an alternative argument
\cite{TS2002b,TS2003} for the self-gravitating stellar system, 
where the modified Clausius relation
\cite{AMPP2001,Abe2001,MPP2001} 
\begin{equation}
  \label{eq: modified_clausius}
  dS_q = \frac{1}{\Tphys}\{ 1+(1-q)\,S_q\}\,d'Q. 
\nonumber
\end{equation}
plays a significant role.

From the expression for the normalization factor $\Zq$ of 
the escort distribution \eqref{eq: escort_dist}, 
the Tsallis entropy 
\eqref{eq: Tsallis entropy}  of the extremum state is given by  
\begin{equation} 
S_q = \dfrac{1}{1 - q} \,\,(\Zq-1).
\nonumber
\label{eq: S_q_extreme}
\end{equation} 
Let us substitute this expression into \eqref{eq: def_Tphys}. 
Then, the relation \eqref{eq: beta-P_relation} for $\Zq/\beta$  
gives us the expression of the thermodynamic temperature $\Tphys$ 
as an integral of a physical quantity, i.e. the pressure distribution:
\begin{equation}
\Tphys = \dfrac{\Zq}{\beta} = \int d^3x\,P(x). 
\end{equation}
We can evaluate the above integral to represent the thermodynamic 
temperature in terms of 
the pressure $P_e$, the density $\rho_e$ at the boundary  and 
the total mass $M$ as 
\begin{equation}
\Tphys  = - \frac{1}{n-5}
\left\{8\pi\,r_e^3P_e-(n+1)\frac{MP_e}{\rho_e}+\frac{GM^2}{r_e}\right\}
\label{eq: thermo_tempeature}
\end{equation}
from which  the dimensionless inverse temperature is denoted by  
  \begin{equation}
\label{eq: eta}
\eta\equiv\frac{GM^2}{r_e\Tphys} = \frac{(n-5)\,\ve}{n+1-2\ue-\ve}    
  \end{equation}
as a function of the homology invariant at the boundary 
\cite{TS2003}.
From \eqref{eq: energy_uv} and \eqref{eq: eta}, the specific heat at 
constant volume $\Cv$ is given by 
\begin{equation}
  \label{eq: specific_heat}
  \Cv \equiv \left(\frac{d \,E~~~}{d\,\Tphys}\right)_e
  = \eta^2 \frac{\displaystyle \left(\frac{d \lambda}{d\xi}\right)_e}
  {\displaystyle \left(\frac{d \eta}{d\xi}\right)_e}. 
\end{equation}

\begin{figure}[t]
\label{fig: Cv_poly}
\begin{minipage}{\textwidth}
   \includegraphics*[width=5.7cm]{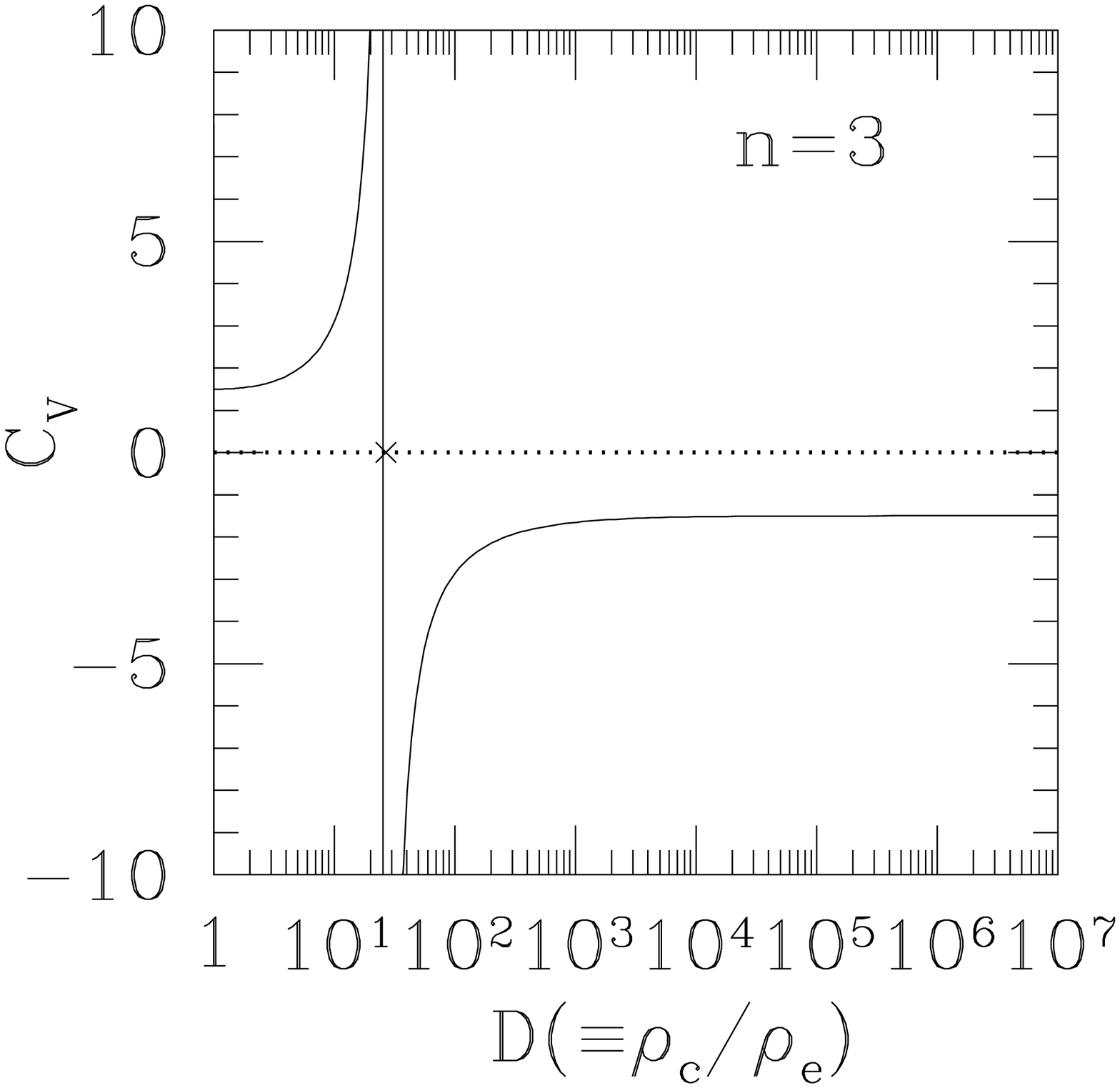}
   \hfill
   \parbox[b]{10.cm}{
   \includegraphics*[width=5.7cm]{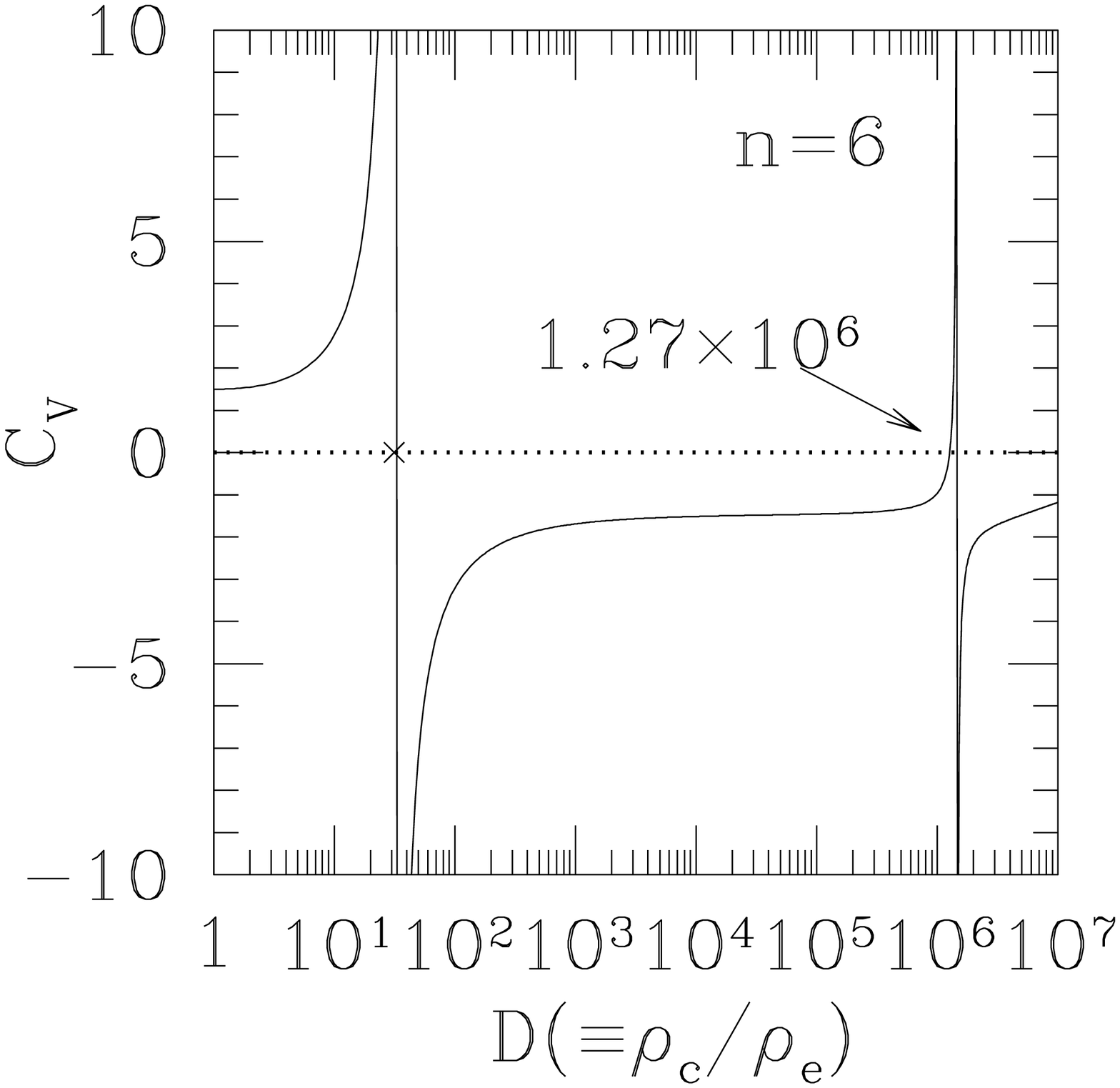}
   \parbox[b]{4cm}{
   \caption{Specific heat as a function of the density contrast for 
            the stellar polytropes with $n=3$ (left) and $n=6$ (right).}
   }
   }
\end{minipage}
\end{figure}%
In order to compare with the behavior of the specific heat 
in the isothermal case of Fig. \ref{fig: Cv_iso}, 
we plot $\Cv$ with respect to the density contrast for several values of 
the polytropic index in Fig. 3.2.

As already denoted at the end of subsection \ref{sec: Gravo_stellar},
the stellar polytropic system exhibits the gravothermal catastrophe
when the polytropic index $n$ is larger than the critical value, 
i.e. $n > 5$. In the isothermal case, the onset of the thermodynamic 
instability is characterized by the condition \eqref{eq: C-outer-inner}. 
Figure \ref{fig: Cv_poly} shows that the thermodynamic argument
based on the negative specific heat for the gravothermal catastrophe 
can be extended to the case of the stellar polytrope, i.e. the equilibrium
of the self-gravitating system described by the extremum of the Tsallis 
entropy.

\begin{figure}
\begin{minipage}{\textwidth}
   \includegraphics*[width=11cm,height=6cm]{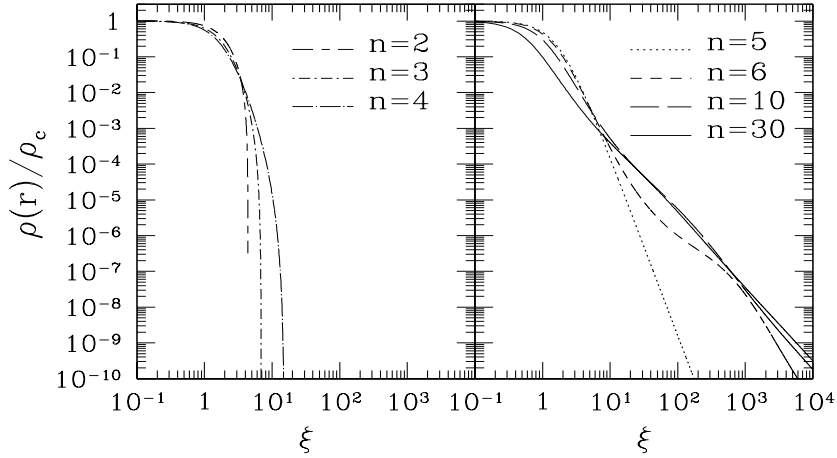}
   \hfill
   \parbox[b]{4cm}
         {\caption[]{Density profiles of the stellar polytropes for 
                      $n < 5$ (left) and $n\geq5$ (right). }
          \label{fig: profile}}
\end{minipage}
\end{figure}%
In Fig. \ref{fig: profile}, we depict the density profiles with respect to
the dimensionless radius $\xi$ \eqref{eq: dimensionless}. 
Clearly, profiles with index $n<5$ rapidly fall off and they 
abruptly terminate at finite radius({\it left panel}), while 
the $n\geq5$ cases infinitely continue to extend over the outer 
radius({\it right panel}). It follows that  stellar polytropic systems with
index $n\geq5$ are able to possess a sufficient amount of the outer normal part 
to realize the gravothermal catastrophe. 

Before closing this section, we mention  differences between 
the old and new Tsallis formalisms. 
First, the relationship between the polytrope 
index $n$ and theTsallis parameter $q$ in (\ref{eq: index}) 
differs from the one obtained previously, but is related to 
it through the {\it duality transformation}, $q\leftrightarrow 1/q$ 
(see (14) in \cite{TS2002a} or (12) in \cite{TS2002b}). 
This property was first addressed in \cite{TMP1998} in a more general context, 
together with changes to the Lagrangian multiplier $\beta$.  
The duality relation implies that all of the thermodynamic properties 
in the new formalism can also be translated into those obtained in the old 
formalism.
Secondly , in previous studies \cite{TS2002b}
based on the standard linear means, the 
radius-mass-temperature relation and the specific heat seriously depend on 
the dimensional parameter for  the phase element $h$ \eqref{eq:measure}. 
By contrast, in the present case \cite{TS2003},  
the radius-mass-temperature relation was 
derived from the non-trivial relation (\ref{eq: beta-P_relation}), 
in which no such $h$-dependence appears. 
The resultant specific heat is also free from this 
dependence, which is a natural outcome of the 
new framework using the normalized $q$-expectation values. Therefore,  
it seems likely that the new formalism provides a better 
characterization for non-extensive quasi-equilibrium systems.

\section{Reality of the stellar polytrope as a quasi-equilibrium state}
\label{sec:4}

\begin{figure}
\begin{minipage}{\textwidth}
   \includegraphics*[width=9cm]{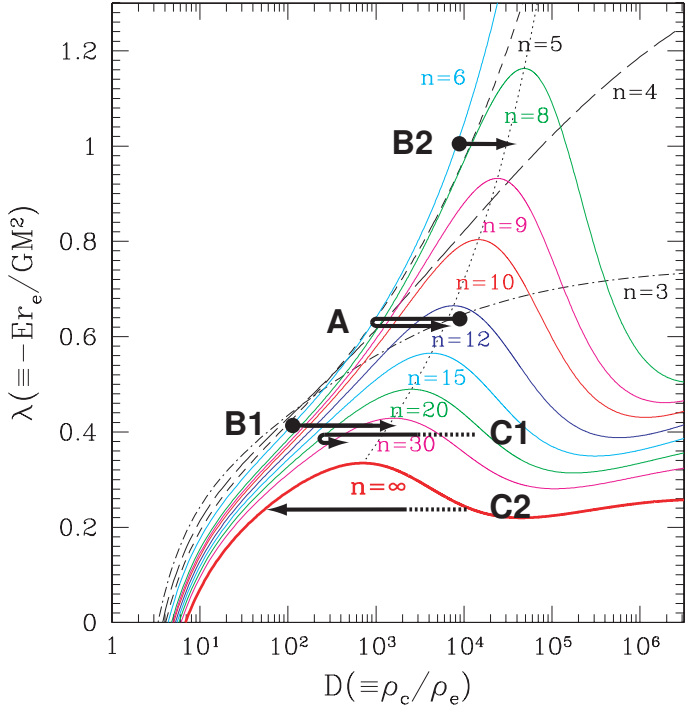}
   \hfill
   \parbox[b]{6cm}
         {\caption[]{Equilibrium sequences of the stellar polytrope 
                     and the isothermal distribution ($n=\infty$) 
                     in the $\lambda\equiv-r_eE/(GM^2)$ vs $ D\equiv\rho_c/\rho_e$ 
		     plane. The thick arrows denote the evolutionary tracks in 
                     each simulation run (see Sec.\ref{sec:4}).}
          \label{fig: Eq_sequence}}
\end{minipage}
\end{figure}

So far, we have discussed the extension of the thermodynamic analysis to
the stellar polytrope which is obtained by applying the variational procedure to 
the Tsallis entropy for the stellar self-gravitating system. 
It has been shown that  thermodynamic quantities 
such as the specific heat are useful for studying the emergence of the 
{\it gravothermal catastrophe} in the stellar polytrope.
In Fig. \ref{fig: Eq_sequence} we summarize our results.
The stellar polytrope confined by an 
adiabatic wall is shown to be thermodynamically stable 
when the polytrope index $n<5$. In other words, if $n>5$, a stable 
equilibrium state ceases to exist for a sufficiently large 
density contrast $D>\Dcrit$,   
where the critical value $\Dcrit$ given by a function of $n$ is determined 
from the second variation of the entropy around the extremum state of the Tsallis 
entropy, $\delta^2S_q=0$\cite{TS2002a,TS2003}. 
The dotted line in Fig.\ref{fig: Eq_sequence} represents 
the critical value $\Dcrit$ for each polytrope index, which indicates that 
the stellar polytrope at low density contrast $D<\Dcrit$ is 
expected to remain stable.

The above arguments indicate that, similar to the isothermal state, 
the stellar polytropic distribution can also be regarded as 
an equilibrium state, since it is described by the extremal state 
of the Tsallis entropy. However, the one-particle distribution function of
the stellar polytrope \eqref{eq: poly_dist} clearly shows that 
the velocity dispersion, 
\begin{equation}
\sigma (r) \propto 
\frac{1}{\rho(r)}\,\,\int\frac{d^3\vv}{h^3} \,\,\, v^2 \,
f(\xx,\vv),
\end{equation}
depends on the radius $r$. 
Only in the isothermal case, $n \rightarrow \infty$, is       
$\sigma$ spatially constant. 
Thus, it is expected that a gradient of the velocity dispersion is relaxed 
within the  time scale due to two-body collisions \eqref{eq:rtime}. 
This means that  the stellar polytrope is no longer the equilibrium but 
a quasi-equilibrium state. In this section, we report the results of  
$N$-body simulations \cite{TSPRL} which were carried 
out to investigate how the stellar polytrope actually evolves.

The $N$-body experiment considered here is the same 
as that investigated in classic papers (\cite{Antonov1962,LW1968}; see 
also  \cite{EFM1997}). 
That is, we confine the 
$N$ particles interacting via Newtonian gravity in a spherical adiabatic wall, 
which reverses the radial components of the velocity if the particles reach 
the wall. Without loss of generality, we set the units as $G=M=\re=1$. 
Note that the typical time scales appearing in this system are 
the dynamical time $t_{\rm dyn}$ \eqref{eq:dtime} and the global relaxation time 
driven by the two-body encounter, $t_{\rm rel}$ \eqref{eq:rtime}, 
which are basically scaled as $t_{\rm dyn}\sim 1$ and 
$t_{\rm rel}\sim 0.1N/\ln N$ in our units. 
%
%
%
%
%
\begin{table*}[b]
\caption{\label{tab:initial_model}Initial distributions and their 
evolutionary states}
\begin{tabular}{cclccc}
\hline
 run no. & initial distribution & parameters & no. of particles  & transient state & final state \\
\hline
 A & stellar polytrope($n=3$) & $D=10,000$  & 2,048 &stellar polytrope& collapse \\
 B1& stellar polytrope($n=6$) & $D=110$     & 2,048 &stellar polytrope& collapse \\ 
 B2& stellar polytrope($n=6$) & $D=10,000$  & 2,048 &stellar polytrope& collapse \\
 C1& Hernquist model          & $a/r_e=0.5$ & 8,192 &stellar polytrope& collapse \\
 C2& Hernquist model          & $a/r_e=1.0$ & 8,192 & none &  isothermal \\
\hline
\end{tabular}
\end{table*}
%
%
%
%
%
Table \ref{tab:initial_model} summarizes 
the five simulation runs. 
Here, in addition to the stellar polytropic  initial state, we also 
consider the non-stellar polytropic state of 
the Hernquist model \cite{H1990}, 
which was originally introduced to account 
for the empirical law of observed elliptical galaxies\cite{BT1987}.

As was expected, the numerical simulations reveal that 
the stellar polytropic distribution gradually changes with time,  
on the time scale of two-body relaxation \eqref{eq:rtime}. 
Furthermore, focusing on the evolutionary 
sequence, we found that the transient state starting 
from the initial stellar polytrope can be remarkably characterized by 
a sequence of stellar polytropes (run A, B1, and B2). 
This is even true in the case starting from the Hernquist model (run C1). 

\begin{figure}[t]
\begin{minipage}{\textwidth}
   \includegraphics*[width=7.7cm]{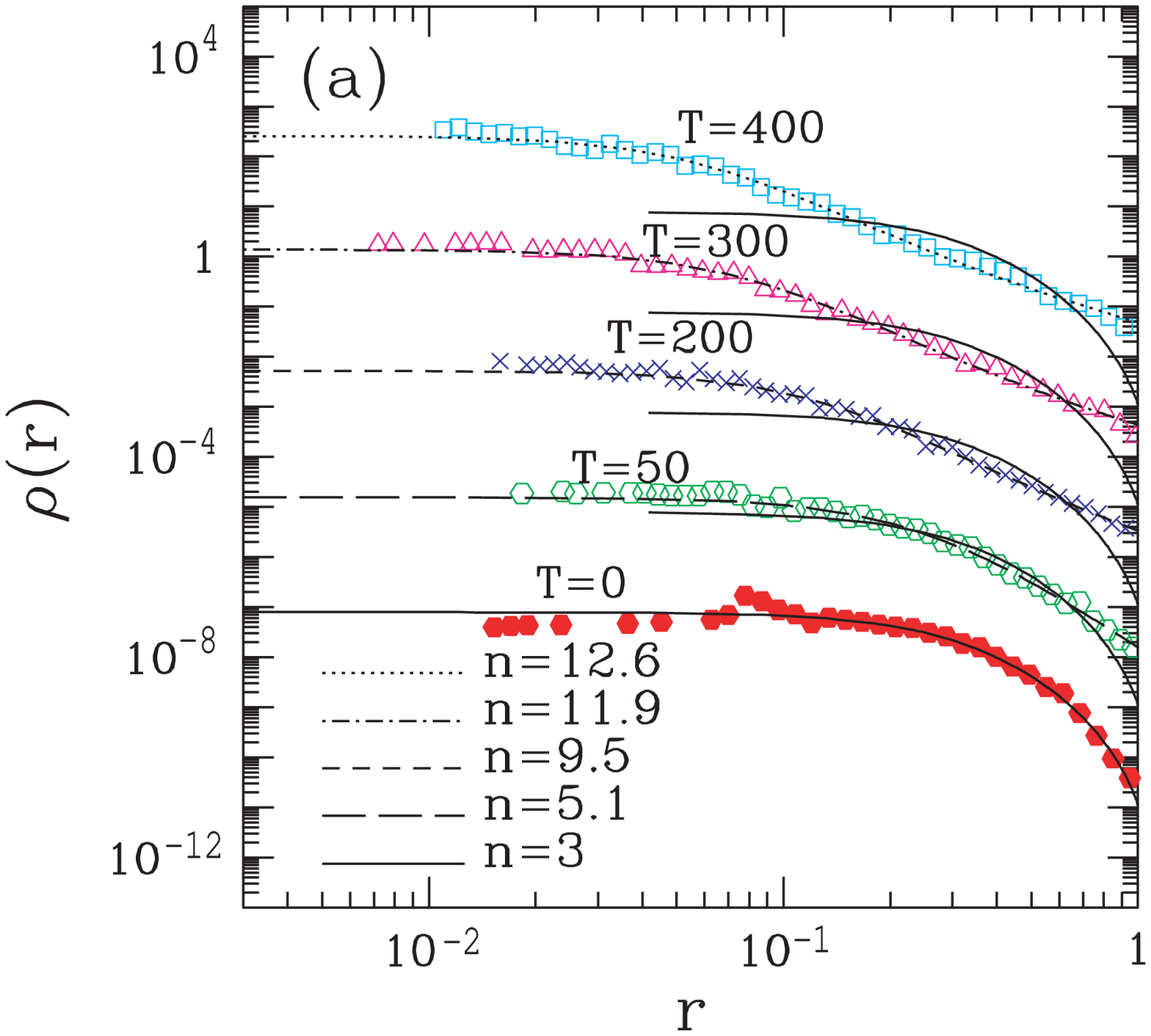}
   \hfill
   \parbox[b]{8.5cm}{
      \parbox[b]{0.5cm}{~~~~~ }
      \includegraphics*[width=7.7cm]{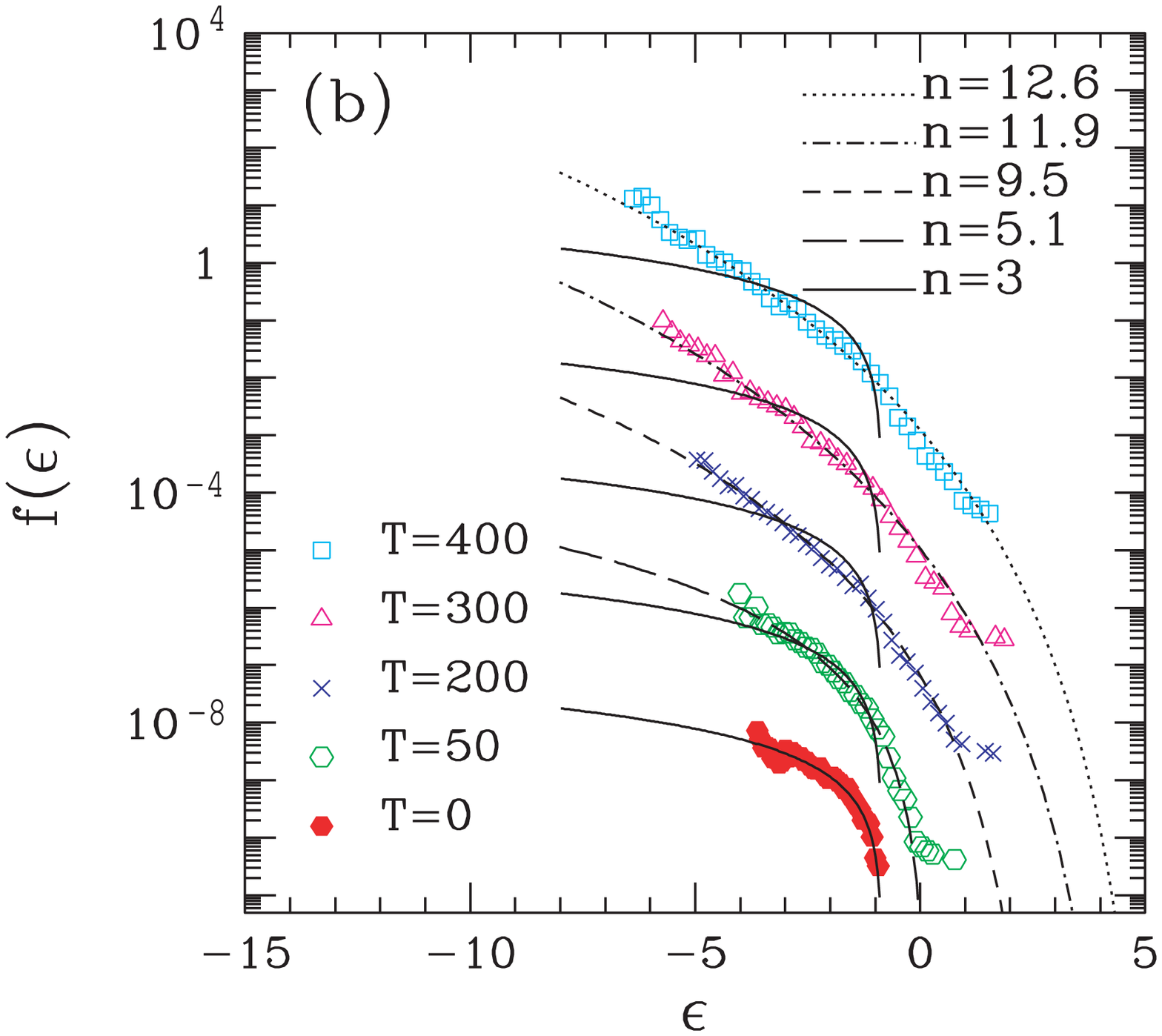}
   }
\end{minipage}
\caption[]{Results from simulation run A: 
(a) snapshots of the density profile $\rho(r)$, and
(b) snapshots of the one-particle distribution function $f(\epsilon)$.}
\label{fig:run_A}
\end{figure}%

Let us show  representative results taken from run A 
(Fig.\ref{fig:run_A}). 
Figure \ref{fig:run_A}(a) plots snapshots of the density profile 
$\rho(r)$, while Fig.\ref{fig:run_A}(b) represents the distribution 
function $f(\varepsilon)$ as a function of the specific energy 
$\varepsilon = \frac{1}{2}v^2 + \Phi(\xx)$. 
Note,  for illustrative purpose, that
each output result is artificially shifted to the two digits 
below. Only the final output with $T=400$ represents the correct 
scales. In each figure, solid lines mean the initial stellar 
polytrope with $n=3$ and the other lines indicate the fitting 
results to the stellar polytrope by varying the polytrope index $n$. 
Note that the number of fitting parameters  
just dexreases to one, i.e. the polytrope index, since the total energy 
is well-conserved in the present situation. Fig.\ref{fig:run_A} 
shows that while the system gradually deviates from the initial 
polytropic state, the transient state 
still follows a sequence of stellar polytropes. 
The fitting results are remarkably good until the time exceeds 
$T\simeq400$, corresponding to $15\, t_{\rm rel}$. After that, 
the system enters the gravothermally 
unstable regime and finally undergoes the core collapse. 

Now, focus on the evolutionary track in each simulation run 
summarized in the energy-radius-density contrast plane (Fig.\ref{fig: Eq_sequence}), 
where the filled circle represents the initial stellar polytrope. 
Interestingly, the density contrast 
of the transient state in run A initially 
decreases, but it eventually turns to increase. 
The turning point roughly corresponds to the stellar polytrope with index 
$n\sim5-6$. Note, however, that the time evolution of the polytrope index itself 
is a monotonically increasing function of time \cite{TSPRL}.
This is indeed true for 
the other cases, indicating the Boltzmann $H$-theorem 
that any  self-gravitating system tends to approach the isothermal state. 
A typical example is  run C2, which finally reaches the 
stable isothermal state. However, as already shown in run A, 
not all the systems can reach the isothermal state. 
Fig.\ref{fig: Eq_sequence} indicates that no isothermal state is possible 
for a fixed value $\lambda>0.335$\cite{Antonov1962,LW1968}, which can be derived from 
the peak value of the trajectory. 
Further, stable stellar polytropes cease to exist at a high density contrast, 
$D>\Dcrit$. In fact, our simulations starting from the stellar polytropes  
finally underwent core collapse due to the gravothermal 
catastrophe (runs A, B1, and B2). 
Though it might not be rigorously correct, the predicted value $\Dcrit$ 
provides a crude approximation to the boundary between stability 
and instability.

\begin{figure}[t]
\begin{minipage}{\textwidth}
   \includegraphics*[width=7.7cm]{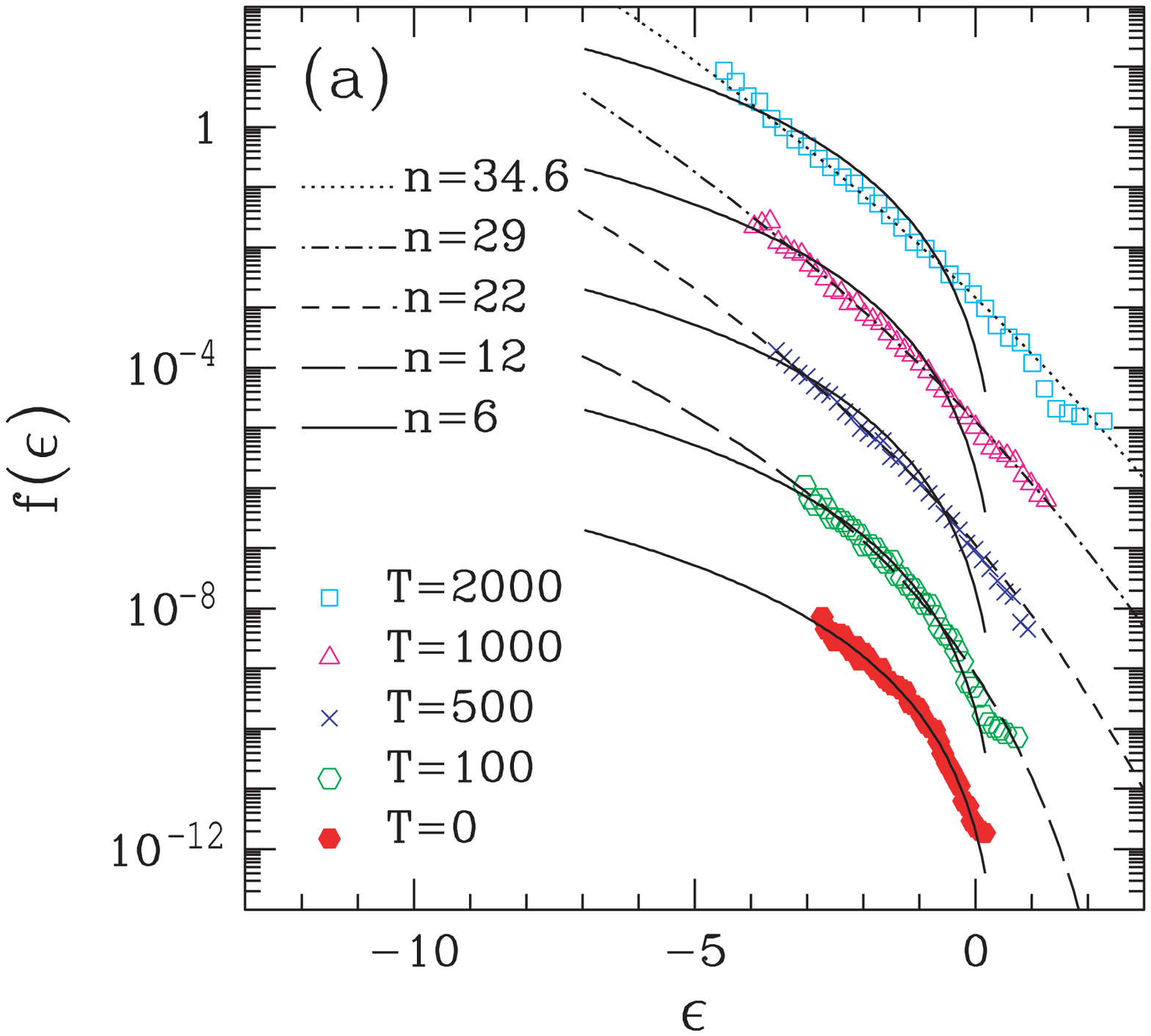}
   \hfill
   \parbox[b]{8.5cm}{
      \parbox[b]{0.5cm}{~~~~~ }
      \includegraphics*[width=7.7cm]{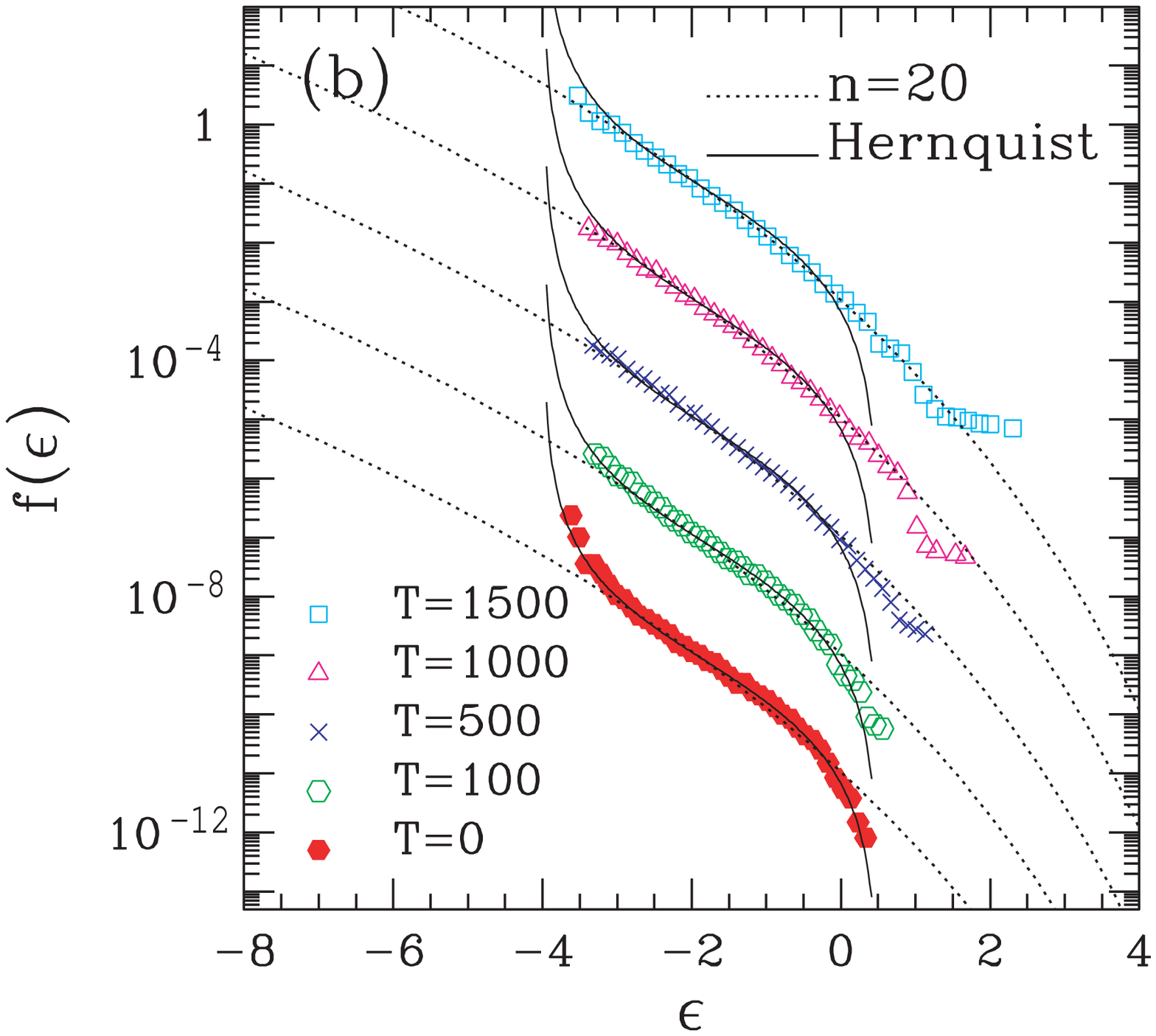}
   }
\end{minipage}
\caption[]{Evolution of the one-particle distribution function in
other models: (a) run B1; (b) run C1}
\label{fig:fe_others}
\end{figure}%

Fig.\ref{fig:fe_others} plots  snapshots of the distribution function taken 
from the other runs. The initial density contrast in run B1 
(Fig.\ref{fig:fe_others}(a)) is relatively low ($D=110$) and therefore the system 
slowly evolves by
following a sequence of stellar polytropes. After $T=2000\sim 74\,\, t_{\rm rel}$, 
the system begins to deviate from the stable equilibrium sequence, 
leading to  core collapse. Another noticeable case is  run C1 
(Fig.\ref{fig:fe_others}(b)). 
The Hernquist model as the initial distribution of run C has a cuspy density 
profile, $\rho(r)\propto 1/r/(r+a)^3$, which behaves as $\rho\propto r^{-1}$ 
at the inner part 
\cite{H1990}. The resultant distribution function $f(\varepsilon)$ shows  
singular behavior at the negative energy region, which cannot be described by the 
power-law distribution. 
Soon after, however, gravothermal expansion\cite{EFM1997}  
takes place and a 
flatter core is eventually formed. Then the system settles into a sequence 
of stellar polytropes and can be approximately described by the stellar 
polytrope with index $n=20$ for a long time. 
Thus the stellar polytrope can be regarded as a quasi-attractor and a 
quasi-equilibrium state. 
%
%
%
%
%
%
\section{Conclusion and Discussion}
%
%
%
%
%
%
%
In this paper, we discussed issues arising from the Tsallis entropy for
the thermodynamic properties of 
stellar self-gravitating systems, 
with a particular emphasis on the standard framework using 
normalized $q$-expectation values. It  turns out that 
the new extremum-entropy state essentially remains unchanged from  
previous studies and is characterized by the 
stellar polytrope, although the distribution function shows several 
distinct properties. By considering these facts carefully, 
the thermodynamic temperature of the extremum state was identified 
through the modified Clausius relation and the specific heat was 
evaluated explicitly. 
A detailed analysis of the behavior of 
specific heat finally led to the conclusion that  
the onset of gravothermal instability remains unchanged with respect to
choice of the statistical average 
for a system confined by an adiabatic wall (micro-canonical case).  
As for a system surrounded by a thermal wall (canonical case),
although the analysis has been skipped in this article,
the stability of the system drastically depends on the choice of the statistical
average \cite{TS2002b,TS2003}.
The existence of these thermodynamic instabilities can also be deduced 
rigorously from the variation of the entropy and free energy  
\cite{TS2002a,TS2002b,TS2003}.  
As a result, above  certain critical values 
of $\lambda$ or $D$, 
the thermodynamic instability appears at $n>5$ for a system confined by 
an adiabatic wall.

We  performed a set of 
numerical simulations of long-term stellar dynamical evolution 
away from the isothermal state and found that 
the transient state of a system confined by an adiabatic wall 
can be remarkably fitted by a sequence of stellar 
polytropes. This is even true for the case in which the outer boundary
in removed \cite{TSPRL}. 
Therefore, the stellar polytropic distribution can be a quasi-attractor 
and a quasi-equilibrium state of a self-gravitating system.

\section*{Acknowledgments}
We are grateful to P.H. Chavanis for comments and discussions. 
This work was supported by Grants-in-Aid from the Japan Society for 
the Promotion of Science (No. 15540368 and No. 14740157).

\appendix
\section{Appendix:~~Evaluation of the relaxation time}
Here we briefly evaluate the relaxation time for
a self-gravitating many-body system \eqref{eq:rtime}. 
For a more precise derivation, consult \cite{BT1987,Chandra}.
In the kinetic theory, a time scale of relaxation due to two-body 
collisions is given by
\begin{equation}
t_{\rm rel} \sim \dfrac{1}{\sigma n v},
\label{ap:rtime}
\end{equation}
where $n$ is the mean number density and $v$ the average relative velocity of
particles. In order to evaluate theamplitude of a two-body collision by 
the mutual gravitational force, we introduce the gravitational radius $r_{\rm g}$
as follows,
\begin{equation}
m v^2 \sim \frac{Gm^2}{r_{\rm g}}~~~~\longrightarrow ~~~~ 
r_{\rm g} \sim \frac{Gm}{v^2}.
\end{equation}
If the impact parameter of the collision between two particles of identical 
mass $m$ becomes smaller than the gravitational radius, i.e. $b < r_{\rm g}$,
orbits of particles are significantly changed by the close encounter.
It follows that the cross section of a two-body collision is given by
\begin{equation}
\sigma \sim 4\pi r_g^2 . 
\end{equation}
By substituting this estimate and $n \sim N/R^3$ into \eqref{ap:rtime}, 
we obtain
\begin{equation}
t_{\rm rel} \sim \frac{v^3 R^3}{4\pi\, G^2 \,m^2 \, N \, 
\ln \left( R/r_g \right) },
\label{ap:rtime2}
\end{equation}  
where $R$ and $N$ are the system size and the number of particles, 
respectively. 
In the above expression \eqref{ap:rtime2}, we have included the  
so--called {\it Coulomb logarithm} term, $\ln (R/r_g)$, which appears due to 
the long-range nature of the interaction \cite{BT1987} and is well-known 
in plasma theory. By taking its ratio to the dynamical time 
$t_{\rm dyn} \sim R/v$, we obtain 
\begin{equation}
{t_{\rm rel}} \sim \frac{N}{4\pi \ln N}\, {t_{\rm dyn}}\,,
\end{equation}
where the average velocity is estimated as $v^2 \sim {GNm}/{R}$ 
from the virial theorem \eqref{eq:virial}.
%
%
%
%
%
%
%

\end{document}